\documentclass[%
 reprint,
superscriptaddress,
nofootinbib,
 amsmath,amssymb,
 aps,
]{revtex4-2}
\usepackage{mathtools}
\usepackage{aps_macros}
\usepackage{pifont}
\usepackage{graphicx}% Include figure files
\usepackage{dcolumn}% Align table columns on decimal point
\usepackage{bm}% bold math
\usepackage{hyperref}% add hypertext capabilities
\newcommand{\fd}[1]{\textcolor{black}{#1}}
\usepackage{subfigure}%
\usepackage{float}% 
\hypersetup{
     colorlinks   = true,
     citecolor    = blue
}

\hypersetup{colorlinks=true, citecolor=blue, linkcolor = magenta}

\begin{document}

\preprint{APS/123-QED}

\title{\fd{Slowly Rotating Anisotropic Neutron Stars with a Parametrized Equation of State}}

\author{L.~M.~Becerra}
\email{laura.becerra@umayor.cl}
\affiliation{Centro Multidisciplinario de F\'isica, Vicerrector\'ia de Investigaci\'on, Universidad Mayor,  Santiago de Chile 8580745, Chile}
 
\author{E.~A.~Becerra-Vergara}
\email{eduar.becerra@correo.uis.edu.co}
\affiliation{Grupo de Investigaci\'on en Relatividad y Gravitaci\'on, Escuela de F\'isica, Universidad Industrial de Santander A. A. 678, Bucaramanga 680002, Colombia}

\author{F.~D.~Lora-Clavijo} 
\email{fadulora@uis.edu.co}
\affiliation{Grupo de Investigaci\'on en Relatividad y Gravitaci\'on, Escuela de F\'isica, Universidad Industrial de Santander A. A. 678, Bucaramanga 680002, Colombia}

\date{\today}

%%%%%%%%%%%%%%%%%%%%%%%%%%%%%%%%%%%%%%%%%%%%%%%%%%%%%%%%%%%%%%%%%%%%%%%%%%%%%%%%%%%%%%%%%%%
\begin{abstract}

In this work, we study the impact of anisotropy on slowly rotating neutron stars by extending the Hartle–Thorne formalism in general relativity to include anisotropy in pressure up to second order in the angular velocity. We assess the presence of anisotropy within the star by employing a quasi-local relationship.  Our results show that the ratio between the gravitational mass of the fastest anisotropic rotating configurations and the corresponding non-rotating ones ranges from 1.12 to 1.25, consistent with recent findings.  We develop universal relations for the moment of inertia, binding energy, and quadrupole moment of the rotating stars. These relations are tested against various equations of state, which were modeled by a piecewise polytropic function with continuous sound speed.
\end{abstract}
%%%%%%%%%%%%%%%%%%%%%%%%%%%%%%%%%%%%%%%%%%%%%%%%%%%%%%%%%%%%%%%%%%%%%%%%%%%%%%%%%%%%%%%%%%%

                              %display desiblack
\maketitle

%\tableofcontents

%%%%%%%%%%%%%%%%%%%%%%%%%%%%%%%%%%%%%%%%%%%%%%%%%%%%%%%%%%%%%%%%%%%%%%%%%%%%%%%%%%%%%%%%%%%
\section{\label{sec:intro} Introduction}
%%%%%%%%%%%%%%%%%%%%%%%%%%%%%%%%%%%%%%%%%%%%%%%%%%%%%%%%%%%%%%%%%%%%%%%%%%%%%%%%%%%%%%%%%%%

Neutron stars (NS) are remarkable astrophysical objects that are natural laboratories for studying matter under extreme conditions. Although they are commonly modeled as static compact objects \citep{2004Sci...304..536L,2012ARNPS..62..485L}, 
in reality, all NS are rotating. Including rotation in the NS models is essential as it causes the stars to become oblate, breaking their spherical symmetry and leading to significant variations in their properties \citep{2000ApJ...528L..29B,1991ApJ...373..579W,1999ApJ...512..282L,2016ApJ...832...28G,2017LRR....20....7P}. 

Observations have revealed that NSs rotate slowly, with spin frequencies varying from $\sim 0.01$ Hz up to \textcolor{black}{$\sim 716$}~ Hz \citep{caleb2022discovery}. The first-millisecond pulsar discoveblack, PSR 1937+214, has a frequency of $641$ Hz  \citep{backer1982millisecond}, while the NS with the fastest known spin reaches $716$ Hz (PSR J1748-2446ad) \citep{2006Sci...311.1901H}. Slowly rotating NSs include pulsars in NS binary systems, such as PSR J0737-3039A \citep{burgay2003increased} and PSR J1946+2052 \citep{stovall2018palfa}, which are capable of merging within a Hubble time. Both will have dimensionless spins\footnote{\textcolor{black}{ The dimensionless spin is defided as $c J/(G M^2)$, being $J$  the star's angular momentum and $M$ its mass}} of approximately $0.04$ and $0.05$, respectively, at the time of the merger. Some X-ray pulsars have been observed to rotate at slow rates, with frequencies as low as approximately $\sim 10^{-5}$ Hz \citep{sidoli2017ax}.
These observations demonstrate that the slow rotation approximation accurately models the observed NS, \textcolor{black}{ as \cite{2005MNRAS.358..923B} shows it is safely applicable to stars rotating with an angular velocity of up to approximately $6800 \, \sqrt{ (M/1.4 M_\odot)(10 \, {\rm km}/R)^3}$~ s$^{-1}$.}

Traditionally, NSs have been modeled as isotropic entities in hydrostatic equilibrium \citep{2000ARNPS..50..481H,2016ARA&A..54..401O,2016ApJ...820...28O}. However, recent advancements suggest that anisotropy, where pressure varies in different directions, can have significant effects on the structure, stability and observable properties of these stars \citep{2019PhRvD.100j3006B,2021MPLA...3650028B,2023EPJC...83..307B,PhysRevD.109.043025}, particularly those that are rotating \citep{2024PhRvD.110b4052B,2022MPLA...3750188P,2016CQGra..33i5005Y,2022hxga.book...30N}. Anisotropy in NS can arise from several physical mechanisms, including strong magnetic fields \citep{2012MNRAS.427.3406F,2015MNRAS.447.3278B}, relativistic nuclear interactions \citep{1975ARA&A..13..335C}, phase transitions in superfluidity \citep{sokolov1980phase}, pion condensation \citep{1972PhRvL..29..382S}, and other exotic phases of matter \citep{1972ARA&A..10..427R, 1997PhR...286...53H,2003GReGr..35.1435D}. These anisotropic pressures can alter the equilibrium star's configuration, influencing key observable properties such as the mass-radius ratio \citep{1974ApJ...188..657B,2021ApJ...922..149D}, the moment of inertia \citep{2020EPJC...80..769R}, maximum mass \citep{2000astro.ph.12265D,2021Ap&SS.366....9R}, tidal deformability \cite{2021EPJC...81..698P,2019PhRvC.100e5804R,2019PhRvD..99j4002B}, non-radial oscillation \citep{2012PhRvD..85l4023D}, blackshift \citep{2021arXiv211212518K} and quadrupole moment \citep{2015CQGra..32n5008S,2015PhRvD..91l3008Y,2021PhRvC.104f5805R,2022PhRvD.106j3518D}. Understanding the impact of anisotropy inside the stars, specially when accounting for rotational effects, is crucial to improving our comprehension of their internal structure and observable properties. In this context, the influence of anisotropy can be constrained by the analysis of observational data. For instance, \citet{2015CQGra..32n5008S} proposed that binary pulsar observations could be used to constrain the degree of anisotropy. Additionally, \citet{2019PhRvD..99j4002B}, \citet{2022PhRvD.106j3518D} and \citet{2023arXiv230515724R} used gravitational wave observations data from GW170817 and GW190814 events to constrain the range of pressure anisotropy within stars. Moreover, the mass and radius measurements obtained from the NICER observations of the pulsars PSR J0030+045 \cite{2019ApJ...887L..21R,2019ApJ...887L..24M} and PSR J0740+662 \cite{2021ApJ...918L..27R,2021ApJ...918L..28M} could provide tighter constraints on the permissible anisotropy range within a NS.

This work explores  the implications of anisotropy in slowly rotating NS, focusing on its impact on macroscopic properties such as mass, radius, angular momentum, moment of inertia, binding energy, and quadrupole moment. We reformulated the structure equations for an anisotropic NS based on the Hartle--Thorne (HT) formalism \citep{1967ApJ...150.1005H,1968ApJ...153..807H}. Following the HT procedure for mass correction and deformation in slowly rotating stars, we determined the Einstein field equations by incorporating anisotropic pressure into the energy-momentum tensor\footnote{\textcolor{black}{While we were working on this paper, \citet{2024PhRvD.110b4052B} published a similar study in which these equations were also derived. Our results agree; however, their study focused on a star with uniform density energy, and  used the Bowers-Liang solution  for the equation of state between the star's radial and tangential pressures. In contrast, this study uses nuclear equations of state modeled by a piecewise polytropic function with a continuous sound speed and  the Horvat relation to describe the anisotropy inside the star.}}. This approach balances computational simplicity with the need to account for the rotation effects on the star’s structure. It allows the study of rotating NS with a particular focus on how anisotropy influences their physical and dynamic properties. Furthermore, we exploblack various nuclear equations of state (EOS) \textcolor{black}{between the energy density and radial pressure} covering different models and matter compositions for NS. We focused solely on physical solutions and compablack our results with the observational constraints. Additionally, we guarantee continuity in radial pressure and sound speed throughout the star using the Generalized Piecewise Polytropic (GPP)  fit  \cite{Boyle2020}. 

To characterize the anisotropy of pressure, we employ the quasi-local relationship suggested by \citet{2011CQGra..28b5009H}, in which the star's compactness and radial pressure are proportional to the difference between the radial and tangential pressure. This \textcolor{black}{EOS between the star's radial and tangetial pressures} ensures that the anisotropy vanishes at the star center,  avoiding singularities in mass and angular momentum \citep{1974ApJ...188..657B}. It is important to mention that, in addition to the Horvat \textit{et al.}  \textcolor{black}{relation}, there are two other well-known models in the literature for pressure anisotropy: \citet{1974ApJ...188..657B} \textcolor{black}{model}, designed to solve the modified TOV equation for an anisotropic star with constant density \citep{1974ApJ...188..657B} and \citet{1981JMP....22..118C} \textcolor{black}{model}, heuristically derived to obtain anisotropic solutions from known solutions for isotropic matter \citep{1981JMP....22..118C}.

The paper structure is as follows. In Sec.~\ref{sec:NEM} we derive the equations of structure for relativistic anisotropic NS up to second order in the angular velocity %using a slow rotation expansion 
 (Sec.~\ref{subsec:TOV}). 
Additionally, we introduce the parametrization used to model various EOS for the NS matter (Sec.~\ref{subsec:EOS}) and provide a brief description of the numerical code employed to integrate the equations  (Sec.~\ref{subsec:NC}). In Sec.~\ref{sec:NSP} we present the macroscopic properties of anisotropic NSs obtained through numerical integration. We analyze angular momentum, mass, and radius for various degrees of anisotropy and compare our results with the observational constraints. Numerical calculations of the universal relations for the moment of inertia, binding energy, and quadrupole moment of the rotating NS are shown in Sec.~\ref{sec:UR}. Finally, we give our concluding remarks in Sec.~\ref{sec:discussion}. Throughout the paper we use geometrical units $(c = G = 1)$, unless explicitly specified otherwise, and signature $(-, +, +, +)$ convention.

%%%%%%%%%%%%%%%%%%%%%%%%%%%%%%%%%%%%%%%%%%%%%%%%%%%%%%%%%%%%%%%%%%%%%%%%%%%%%%%%%%%%%%%%%%%
\section{\label{sec:NEM} Neutron Star in the slow rotation approximation}
%%%%%%%%%%%%%%%%%%%%%%%%%%%%%%%%%%%%%%%%%%%%%%%%%%%%%%%%%%%%%%%%%%%%%%%%%%%%%%%%%%%%%%%%%%%
In this section, we derived the model for a slowly rotating anisotropic NS using the HT formalism. We incorporated anisotropic pressure into the energy-momentum tensor and applied HT's method to solve the Einstein field equations. All parameters describing the star were computed up to second order in angular velocity ($\mathcal{O}(\Omega^2)$).

\subsection{\label{subsec:TOV} Anisotropic Star Structure} 

Consider an anisotropic fluid within a stationary, axially symmetric spacetime. Following the HT formalism \citep{1967ApJ...150.1005H,1968ApJ...153..807H}, the line element can be expressed in terms of the components of the metric tensor $g_{\alpha \beta}$ as follows
\begin{equation}
\begin{aligned}
\mathrm{d}s^2 =& -H^{2} \mathrm{d}\mathrm{t}^{2} + Q^2 \mathrm{d}r^2 \\ 
& + r^2 K^2 \left[\ \mathrm{d}\theta+\sin^2\theta\left(\mathrm{d}\varphi-\omega \mathrm{d}\mathrm{t}\right)^2 \ \right], \label{eq:ds2}
\end{aligned}
\end{equation}
where  $H$, $Q$ and $K$ are %functions of $r$ and $\theta$ 
given by
\begin{eqnarray}
H(r,\theta) &=& e^{\nu(r)/2} \left[\ 1+2  \ h(r,\theta) \ \right]^{1/2}, \label{eq:H} \\ 
Q(r,\theta) &=& e^{\lambda(r)/2} \left[\ 1+ 2 \  \frac{m(r,\theta)}{r}e^{\lambda(r)} \ \right]^{1/2}, \label{eq:Q} \\ 
K(r,\theta) &=& \left[\ 1+ 2 \  k(r,\theta) \ \right]^{1/2}. \label{eq:K}
\end{eqnarray}
%
%being $\xi$ a bookkeeping parameter that counts the order of slow rotation while 
The functions $e^{\nu(r)}$ and $e^{\lambda(r)}$ describe the non-rotating anisotropic star solution of the Tolman-Oppenheimer-Volkoff (TOV) equations (see section~\ref{sec:zeroO}). The parameter $\omega=\omega(r,\theta)$ represents the angular velocity $(d\varphi/d t)$ of the local inertial frame, a quantity directly proportional to the angular velocity of the star, $\Omega$.  The functions $h(r,\theta)$, $m(r,\theta)$, $k(r,\theta)$ and $\omega(r,\theta)$  are perturbative corrections related to the deformation of the star due the rotation and are of order $\Omega^2$:
\begin{eqnarray}
h(r,\theta) &=& h_0(r) + h_2(r) P_2(\cos\theta)+ O(\Omega^4) , \\
m(r,\theta) &=& m_0(r) + m_2(r) P_2(\cos\theta)+O(\Omega^4),\\
k(r,\theta) &=&  k_2(r) P_2(\cos\theta)+O(\Omega^4), \\
\omega(r,\theta) &=&  \omega_1(r) P'_1(\cos\theta)+O(\Omega^3),
\end{eqnarray}
where $P_1$ and $P_2$ are the $l=1$ and $l=2$ Legendre polynomials. We neglect the zero-order term of the  $k(r,\theta)$ function since it can be eliminated by performing a radial coordinate transformation \citep{1967ApJ...150.1005H}. Here $h_0(r)$ and $m_0(r)$ are functions associated with the monopolar deformation while $h_2(r)$, $m_2(r)$ and $k_2(r)$ characterize the radial dependence of the quadrupole deformation.

We  assume that the source of matter is described by an anisotropic perfect fluid whose energy-momentum tensor can be written as~\cite{misner1973,PhysRevD.85.124023}
\begin{eqnarray}\label{eq:Tmunu}
T_{\alpha \beta} &=& (\epsilon + P_\perp) u_\alpha u_\beta + P_\perp g_{\alpha \beta} + (P - P_\perp) n_\alpha n_\beta, \label{T_desc}
\end{eqnarray}
where $P=P(r,\theta)$ is the radial pressure, $P_\perp=P_\perp(r,\theta)$ is the tangential pressure and $\epsilon=\epsilon(r,\theta)$ is the density energy. The vectors $u^{\alpha}$ represent the 4-velocity of the rotating fluid: 
\begin{equation}\label{eq:u_vec}
\begin{aligned}
u^\alpha = & u^t[1,0,0,\Omega ]\\
         = & \ e^{-\nu/2}\left(1-h+ \frac{\bar{\omega}^2}{2} \ e^{-\nu}r^2\sin^2\theta \right)[1,0,0,\Omega],
\end{aligned}
\end{equation}
so satisfies the normalization property $u^\alpha u_\alpha = -1$. The quantity $\bar{\omega}\equiv\Omega-\omega$ is the angular velocity of the fluid relative to the local inertial frame. %and holds significance in the structural equations. 
The vector  $n^{\alpha}$ is a normalized space-like vector, orthogonal to the fluid velocity (i.~e.~ $n^\alpha n_\alpha = 1$ and $u^\alpha n_\alpha =0$). For the anisotropic axial symmetric case, its components are \cite{2024PhRvD.110b4052B}:
\begin{equation}
\begin{aligned}
n^\alpha = & [0 , g_{rr}^{-1/2}, r \mathcal{Y} g_{\theta\theta}^{-1/2}, 0] \\
         = &  \left[0 , \ e^{-\lambda/2}\left(1+ \dfrac{2 \ e^{\lambda}}{r}  \ m\right)^{-1/2}, \frac{\mathcal{Y}}{K}, 0\right],
\end{aligned}
\end{equation}
where  $\mathcal{Y} =\mathcal{Y}(r,\theta) $ is a function of order $\Omega^2$. 

To asure a smooth transition between isotropic and anisotropic regimes, it is essential to introduce a functional form for $P - P_{\perp}$. Following the works \citep{PhysRevD.85.124023,PhysRevD.109.043025}, we define the anisotropy parameter $\sigma \equiv P - P_\perp$, in the non-rotating configuration as 
\begin{equation}\label{eq:Delta_p}
\sigma =  -\lambda_a  \frac{2 M }{r}P \, , 
\end{equation}
 here $\lambda_a$ is a parameter controlling the amount of anisotropy inside the star. %\textcolor{black}{$\lambda_a\approx 1$ when anisotropy arises by pion condensation \citep{PhysRevLett.29.382} while for a NS with maximum mass $\lambda_a\approx-2$ \citep{PhysRevD.85.123004}. In order to encompass the two previous values, and in accordance with references \citep{PhysRevD.109.043025,2011CQGra..28b5009H,PhysRevD.91.123008,PhysRevD.85.124023}, the range of values for the anisotropy parameter used for this work is $-2\le \lambda_a \le 2$.} 
Following \citep{PhysRevD.109.043025}, the range of values for the anisotropy parameter used for this work is $-2\le \lambda_a \le 1$. Since rotation deforms the star and displaces the fluid, energy density and pressures change. Hence, %by adopting the Hartle-Thorne formalism for an anisotropic fluid, 
the variations in pressure, energy density, and anisotropy up to $\mathcal{O}(\Omega^2)$ can be written as follows:
\begin{eqnarray}
     &&\delta P =  P-P_0 = (\epsilon_0 + P_0)\left[\ p_0^\ast + p_2^\ast P_2(\cos\theta)\right]  , \\
    &&\delta\epsilon = \epsilon-\epsilon_0 =  \left.\frac{\mathrm{d}\epsilon}{\mathrm{d}P}\right|_{P_0} (\epsilon_0 + P_0)\left[\ p_0^\ast + p_2^\ast P_2(\cos\theta)\right],  \\
   &&\delta \sigma = \sigma - \sigma_0= 
   \left.\frac{\mathrm{d}\sigma}{\mathrm{d}P}\right|_{P_0} (\epsilon_0 + P_0)\left[\ p_0^\ast + p_2^\ast P_2(\cos\theta)\right],
\end{eqnarray}
where we have consideblack a barotropic EOS, i.e.~$\epsilon = \epsilon(P)$. The functions $\epsilon_0=\epsilon_0(r)$, $P_0=P_0(r)$ and $\sigma_0=\sigma_0(r)$ represent the energy density, pressure, and anisotropic distribution, respectively, computed for the non-rotating anisotropic star. 

% $P_0^{\ast}=P^{\ast}_0(r)$ and $P_2^{\ast}=P_2^{\ast}(r)$ are related with the monopolar and quadrupole deformation of the pressure distribution, respectively.

Using the previous definitions, in the next section, we apply Einstein's field equations,  $G^\mu_\nu= 8\pi T^\mu_\nu$,  to derive first-order differential equations for the functions $m_0(r)$, $p_0^\ast(r)$, $p_2^\ast(r)$, $h_2(r)$, $k_2(r)$, $\mathcal{Y}(r,\theta)$, along with an algebraic equation for $m_2(r)$. 
\subsubsection{Zero and first order equations:}\label{sec:zeroO}
The Einstein equations for the line element given in (\ref{eq:ds2}) and the energy-momentum tensor given in (\ref{eq:Tmunu}) blackuce to the anisotropic TOV equation up to $\mathcal{O}(\Omega^0)$  \cite[see][for details]{PhysRevD.109.043025}:
\begin{eqnarray} 
\dfrac{\mathrm{d}M}{\mathrm{d}r} &=& 4\pi r^{2}\epsilon_0 , \label{eq:TOV_a} \\
\dfrac{\mathrm{d}\nu}{\mathrm{d}r} &=& \dfrac{2\left(4\pi r^3 P_0 + M\right)}{r\left(r-2M\right)}, \label{eq:TOV_b}\\
\dfrac{\mathrm{d}P_0}{\mathrm{d}r} &=& -\dfrac{1}{2}\left(\epsilon_0+P_0\right) \dfrac{\mathrm{d}\nu}{\mathrm{d}r}-\dfrac{2\sigma_0}{r}, \label{eq:hyd}
\end{eqnarray}
with $e^{-\lambda} = 1-2M/r$. The function $\bar{\omega}(r)$, accountable for the dragging of inertial frames, 
is obtained from the equation $G^t_\varphi= 8\pi T^t_\varphi$ up to order $\mathcal{O}(\Omega)$
\begin{equation}\label{eq:omega}
\begin{aligned}
    \dfrac{\mathrm{d}^2\bar{\omega}}{\mathrm{d}r^2} = & \dfrac{4\pi r^2(P_0+\epsilon_0)}{r-2M}\left[\dfrac{4\bar{\omega}}{r}\left(1-\frac{\sigma_0}{P_0+\epsilon_0}\right)+\dfrac{\mathrm{d}\bar{\omega}}{\mathrm{d}r}\right] \\
      & - \dfrac{4}{r}\dfrac{\mathrm{d}\bar{\omega}}{\mathrm{d}r}.
\end{aligned}
\end{equation}

\subsubsection{Monopolar Deformation}
At $\mathcal{O}(\Omega^2)$, with  the $l=0$ part of the  equation $G_t^t= 8\pi T_t^t$ and  $G_r^r= 8\pi T_r^r$,  we can obtain the following expression:
\begin{equation}\label{eq:m0}
\begin{aligned}
     \dfrac{\mathrm{d}m_0}{\mathrm{d}r} = & \ 4\pi r^2\dfrac{\mathrm{d}\epsilon_0}{\mathrm{d}P_0}(P_0+\epsilon_0)p_0^* \\
     & + \dfrac{e^{-\nu}r^3}{12} \left(r-2M\right)\left(\dfrac{\mathrm{d}\bar{\omega}}{\mathrm{d}r}\right)^2 \\
     & + \dfrac{8\pi}{3} e^{-\nu}r^4 \left(\epsilon_0 + P_0-\sigma_0 \right)\overline{\omega}^2\, ,
\end{aligned}
\end{equation}
%
%and combining the hydrostatic equation with the $l=0$ part of the equations $G_r^r= 8\pi T_r^r$, we can obtain the equation for $P_0^\ast(r)$
%
\begin{equation}\label{eq:pstar0}
\begin{aligned}
      \dfrac{\mathrm{d}p_0^\ast}{\mathrm{d}r} = & - \dfrac{m_0\left(1+8\pi r^2P_0\right)}{\left(r-2M\right)^2}-\dfrac{4\pi r^2p_0^\ast \left(\epsilon_0+P_0\right)}{r-2M}\\
      &- \frac{2}{r}\left[\frac{\mathrm{d}\sigma_0}{\mathrm{d}P_0} -\frac{  \sigma_0}{(\epsilon_0+P_0)}\left(1+\dfrac{d\epsilon_0}{dP_0}\right)\right]p_0^\ast \\
      & + \dfrac{r^3}{12}e^{-\nu}\left(\dfrac{\mathrm{d}\bar{\omega}}{\mathrm{d}r}\right)^2 + \frac{r^2}{3}e^{-\nu}\left(1-\dfrac{\sigma_0}{\epsilon_0+P_0}\right)\\
      & \times \left[\left(\dfrac{2}{r}-\dfrac{\mathrm{d}\nu}{\mathrm{d}r}\right) \bar{\omega}^2+ 2 \bar{\omega}\dfrac{\mathrm{d}\bar{\omega}}{\mathrm{d}r}\right].
\end{aligned}
\end{equation}

To write the above equation, we have used the $l=0$ part of the radial component of the  hydrostatic equilibrium equation ($T_{r\nu;\nu}=0$):
\begin{equation}\label{eq:h0}
\begin{aligned}
     \dfrac{\mathrm{d}h_0}{\mathrm{d}r} = & - \dfrac{\mathrm{d}p^\ast_0}{\mathrm{d}r} - \dfrac{2}{r}\frac{\mathrm{d}\sigma_0}{\mathrm{d}P_0}p_0^\ast\\
      & +\frac{2 \sigma_0}{r(\epsilon_0+P_0)}\left(1-\dfrac{d\epsilon_0}{dP_0}\right) p^\ast_0 \\
     & -\dfrac{1}{3}\left(1-\dfrac{\sigma_0}{P_0+\epsilon_0}\right)\dfrac{\mathrm{d}}{\mathrm{d}r}\left(e^{-\nu}\bar{\omega}^2 r^2\right).
\end{aligned}
\end{equation}
When the anisotropic parameter cancels ($\sigma_0=\sigma_1=0$), we recover the structure equations for isotropic slowly rotating configurations as in \cite{1967ApJ...150.1005H}. Specifically, equation~(\ref{eq:h0}) simplifies to an algebraic relation between $p^\ast_0$ and $h_0$: $h_0=-p_0^\ast - \frac{1}{3}r^2\bar{\omega}^2e^{-\nu}+Cte$. This relation only holds in the isotropic case but was erroneously assumed to apply to the anisotropic case in \cite{2021EPJC...81..698P}. It is worth noting that our equations coincide with those in \cite{2024PhRvD.110b4052B}.

%%%%%%%%%
\subsubsection{Quadrupolar Deformation:}
At $\mathcal{O}(\Omega^2)$,  we can obtain an algebraic relation for $m_2(r)$ using $R^\theta_\theta-R^\varphi_\varphi=8\pi(T^\theta_\theta-T^\varphi_\varphi)$:
\begin{equation}\label{eq:m2}
\begin{aligned}
     \frac{m_2}{r-2M} = & - h_2 + \dfrac{8\pi}{3}r^4 e^{-\nu} \left(\epsilon_0+P_0-\sigma_0\right) \bar{\omega}^2\\
      & +  \dfrac{1}{6}r^3 e^{-\nu} \left(r-2M\right)\left(\dfrac{\mathrm{d}\bar{\omega}}{\mathrm{d}r}\right)^2.
\end{aligned}
\end{equation}
The $l=2$ part of the equations $G^r_r=8\pi T^r_r$, $R^\theta_r=0$, and $T_{r\nu;\nu}=0$ gives differential equations for $h_2(r)$, $v_2(r)= k_2(r)+h_2(r)$ and $p_2^\ast(r)$ \cite{2015PhRvD..91l3008Y,2024PhRvD.110b4052B}:
\begin{equation}\label{eq:v2}
\begin{aligned}
      \dfrac{\mathrm{d}v_2}{\mathrm{d}r} = & - h_2\dfrac{\mathrm{d}\nu}{\mathrm{d}r}+\dfrac{1}{12}e^{-\nu}\left(2+r\dfrac{\mathrm{d}\nu}{\mathrm{d}r}\right)\left[r^2\left(r-2M\right)  \left(\dfrac{\mathrm{d}\bar{\omega}}{\mathrm{d}r}\right)^2\right.\\
      & \left.  + \dfrac{16\pi\bar{\omega}^2r^3}{\left(\epsilon_0+P_0-\sigma_0\right)^{-1}} \right] +\frac{8\pi}{3}r^2\sqrt{\frac{r}{r-2M}}Y,
\end{aligned}
\end{equation}
\begin{equation}\label{eq:h2}
\begin{aligned}
      \dfrac{\mathrm{d}h_2}{\mathrm{d}r} & = -\dfrac{4h_2M+8\pi r^3p^*_2\left(\epsilon_0+P_0\right)}{r^2\left(r-2M\right)\nu'} \\
      & - h_2\nu' - \dfrac{e^{-\nu}}{6}\left[\dfrac{r^2}{\nu'}-\dfrac{r^3}{2}\left(r-2M\right)\nu'\right] \left(\dfrac{\mathrm{d}\bar{\omega}}{\mathrm{d}r}\right)^2 \\
      & + \dfrac{4\pi r^4 e^{-\nu}\nu'\bar{\omega}^2}{3\left(\epsilon_0+P_0-\sigma_0\right)^{-1}} - \dfrac{2v_2}{M+4\pi r^3P_0}\\
      & +\frac{8\pi }{3}r^2\left(1-\frac{2M}{r}\right)^{-1/2}
      \left(\frac{2}{r\nu'}+1\right)Y,
\end{aligned}
\end{equation}
\begin{equation}\label{eq:Pstar2}
\begin{aligned}
  \dfrac{\mathrm{d}p^\ast_2}{\mathrm{d}r}  = &- \frac{2}{r}\left[\frac{\mathrm{d}\sigma_0}{\mathrm{d}P_0} -\frac{  \sigma_0}{(\epsilon_0+P_0)}\left(1+\dfrac{d\epsilon_0}{dP_0}\right)\right]p_2^\ast\\
     & -  \dfrac{\mathrm{d}h_2}{\mathrm{d}r} +\frac{2\sigma_0}{\epsilon_0+P_0}\left(\dfrac{\mathrm{d}h_2}{\mathrm{d}r}-\dfrac{\mathrm{d}v_2}{\mathrm{d}r}\right)\\ 
     &- \dfrac{1}{3}\left(1-\dfrac{\sigma_0}{P_0+\epsilon_0}\right)\dfrac{\mathrm{d}}{\mathrm{d}r}\left(e^{-\nu}\bar{\omega}^2 r^2\right)\\
      & -\left(1-\frac{2M}{r}\right)^{-1/2}
      \frac{2Y}{P_0+\epsilon_0},
\end{aligned}
\end{equation}
%

%
%Equations (\Ref{eq:h2}) and (\Ref{eq:h2}) are integrated with the boundary conditions $h_2(r=0)=0$ and $v_2(r=0)=0$. Function $m_2(r)$ is obtained from the $l=2$ part of the equation $R^\theta_\theta-R^\varphi_\varphi=8\pi(T^\theta_\theta-T^\varphi_\varphi)$

where $Y(r) = \sigma_0 \mathcal{Y}/(\sin\theta\cos\theta)$. Finally,   equation $T_{\theta\nu;\nu}=0$ gives :
\begin{equation}\label{eq:sigma2}
\begin{aligned}
\sigma_{2} & = p_2^\ast (\epsilon_0+P_0) - \dfrac{m_2 \sigma_0 }{r-2M}\\ 
       &  + \left(\epsilon_0+P_0-\sigma_0\right)(h_2+\dfrac{1}{3}r^2 e^{-\nu} \bar{\omega}^2)\\
       & -\frac{r^2}{3}\left(1-\frac{2M}{r}\right)^{1/2}\left[Y\left(\frac{4}{r}+\frac{\nu'}{2}\right)+\frac{dY}{dr}\right].
\end{aligned}
\end{equation}
 Thus, the system of equations that describe the properties and structure of a slowly rotating anisotropic star up to order $O(\Omega^2)$ is formed by nine first-order differential equations: Eqs.~(\ref{eq:TOV_a})-(\ref{eq:pstar0}) and (\ref{eq:v2})-(\ref{eq:sigma2}), along with one algebraic equation: Eq.~(\ref{eq:m2}). For isotropic configuration, equation~(\ref{eq:sigma2}) is equivalent to equation~(\ref{eq:Pstar2}).
%###################################################
\subsection{\label{subsec:EOS} Equation of State}

We explore the broad range of nuclear EOS from different models and compositions used in \cite{PhysRevD.109.043025} (see Appendix A). 
 The NS EOS have been parameterized using the GPP fit \cite{Boyle2020}. For the mass density, $\rho$, in the range from $\rho_i$ to $\rho_{i+1}$, the pressure $P$ and energy density $\epsilon$ are determined as follows:
\begin{eqnarray}\label{eq:poly_eos}
    P(\rho) &=& K_i\rho^{\Gamma_i} + \Lambda_i,\\
    \epsilon(\rho) &=& \frac{K_i}{\Gamma_i-1}\rho^{\Gamma_i} +(1+a_i)\rho - \Lambda_i\, ,
\end{eqnarray}
where the parameters $K_i$, $\Lambda_i$ and $a_i$ are established by ensuring the continuity and smoothness of the energy density and pressure at the boundaries between density intervals:
\begin{eqnarray}
    K_{i+1}&=& K_i\frac{\Gamma_i}{\Gamma_{i+1}}\rho^{\Gamma_i-\Gamma_{i+1}} ,\label{eq:Ki}\\
    \Lambda_{i+1}&=& \Lambda_i + \left(1-\frac{\Gamma_i}{\Gamma_{i+1}}\right)K_i\rho^{\Gamma_i},\label{eq:Lambdai}\\
    a_{i+1}&=& a_i +\Gamma_i\frac{\Gamma_{i+1}-\Gamma_i}{(\Gamma_{i+1}-1)(\Gamma_i)-1}K_i\rho^{\Gamma_i-1}.\label{eq:ai}
\end{eqnarray}
This method of parameterizing the pressure ensures the continuity of the sound speed within the configuration. In modeling the high-density region of the NS  EOS (densities surpassing the nuclear saturation density, $\rho_s\approx 2.4 \times 10^{14}$~g cm$^{-3}$), we implement a three-zone GPP model  \cite[see Table II of][for the values of the polytropic indices, $\Gamma_i$, and the dividing densities, $\rho_i$]{PhysRevD.109.043025}, while for the low-density region($\rho<\rho_s$), we have utilized the parameterization outlined in \cite{Boyle2020}.
%

%%%%%%%%%%%%%%%%%%%%%%%%%%%%%%%%%%%%%%%%%%%%%%%%%%%%%%%%%%%%%%%%%%%%%%%%%%%%%%%%%%%%%%%%%%%
\subsection{\label{subsec:NC} Numerical Calculations}
%%%%%%%%%%%%%%%%%%%%%%%%%%%%%%%%%%%%%%%%%%%%%%%%%%%%%%%%%%%%%%%%%%%%%%%%%%%%%%%%%%%%%%%%%%%

All the anisotropic slowly rotating NS configurations are computed using a fourth-order Runge-Kutta integrator in a 1D spherical grid extending from $r = 0M$ to the outer domain boundary, $r_{\rm max} = 100 M$.  To avoid the singular behavior at $r=0$, we find series solutions near the origin of the field equations (Eqs.~\ref{eq:TOV_a}-\ref{eq:pstar0},  \ref{eq:v2}-\ref{eq:sigma2}, with Eq.~\ref{eq:m2}), and evaluate them at the  first mesh point located at $r = \Delta r$, being $\Delta r$ the uniform spatial resolution of the grid:
\begin{align}
    M = & \frac{4\pi}{3}  \epsilon_c (\Delta r)^3, \\
    \nu = & \nu_c + 4\pi \left(P_c + \frac{\epsilon_c}{3}\right)(\Delta r)^2 , \\
     P_0 = & P_c - \left[2\pi\left(P_c+\frac{\epsilon_c}{3}\right)\left(P_c+\epsilon_c\right) + \sigma^{(2)}_0\right](\Delta r)^2, \\
     \bar{\omega} = & \bar{\omega}_c +\frac{8\pi}{3}\left(P_c+\epsilon_c\right)\bar{\omega}_c(\Delta r)^2, \\
     m_0 = & \frac{4\pi}{15}(P_c+\epsilon_c) e^{-\nu_c}\bar{\omega}_c^2\left[ \frac{\left(d\epsilon_0/dP_0\right)_c }{1-\sigma_1^{(0)}}  +2\right] (\Delta r)^5, \\
     p_0^{\ast} = & \frac{1}{3}\ \frac{e^{-\nu_c}\bar{\omega}_c^2}{1-\sigma_1^{(0)}} \ (\Delta r)^2, \\
     Y = & \frac{\sigma_1^{(0)}(P_0+\epsilon_c)}{4+\sigma_1^{(0)}}\left(3A+e^{-\nu_c}\bar{\omega}_c^2\right) \Delta r, \\
     v_2 = & -\frac{2\pi}{3}\left[ A\left(3P_c + \epsilon_c \right)-  Y(1)\right.  \nonumber \\
     & \left.-\left(P_c + \epsilon_c \right) e^{-\nu_c}\bar{\omega}_c^2  \right](\Delta r)^4,  \\
     p_2^{\ast} = & \frac{-4}{4+\sigma_1^{(0)}}\left( A+\frac{1}{3}e^{-\nu_c}\bar{\omega}_c^2 \right) (\Delta r)^2, \\
     h_2 = & A (\Delta r)^2 ,   
\end{align}
with
\begin{align}
    \sigma_0^{(2)}  &= -\frac{8\pi}{3}\lambda_a \epsilon_c P_c, \\
    \sigma_1^{(0)} &= \frac{4\lambda_a \epsilon_c P_c}{3P_c^2 -4 (\lambda_a-1)P_c\epsilon_c +\epsilon_c^2} \, ,
\end{align}
for the anisotropy given by (\ref{eq:Delta_p}), and $A$ is a free parameter. In the above equations, it has been assumed that the central values of functions ${\nu, P,\bar{\omega}}$ are ${\nu_c,P_c,\bar{\omega}_c}$, respectively,  while the rest of the functions vanish at the star's center.
Equations (\ref{eq:h2})-(\ref{eq:pstar0}), and (\ref{eq:v2})-(\ref{eq:sigma2}) are integrated up to $R$, the radius of the non-rotating configuration, defined as the one where $\rho(R) \approx 10^{4}$~ g\, cm$^{-3}$. In this sense, the mass of the non-rotating configuration is $M(R)$.

%\textcolor{black}{  In particular, this code has been used to constrain several configurations of neutron and quark stars using the GW170817 observation \cite{2020Ap&SS.365...43A}. In addition, we have used this code to  study anisotropic quark stars with an interacting quark equation of state, where the contribution of the fourth-order corrections parameter a4 of the QCD perturbation on the radial and tangential pressure generates significant effects on the mass-radius relation and the stability of the quark star \cite{2019PhRvD.100j3006B}.}

%%%%%%%%%%%%%%%%%%%%%%%%%%%%%%%%%%%%%%%%%%%%%%%%%%%%%%%%%%%%%%%%%%%%%%%%%%%%%%%%%%%%%%%%%%%
\section{\label{sec:NSP} Neutron Stars Properties}
%%%%%%%%%%%%%%%%%%%%%%%%%%%%%%%%%%%%%%%%%%%%%%%%%%%%%%%%%%%%%%%%%%%%%%%%%%%%%%%%%%%%%%%%%%%

\subsection{\label{subsec:JMI} Angular momentum, mass and radius }
%%%%%%%%%%%
\begin{figure*}
    \centering
    \includegraphics[width=0.99\textwidth]{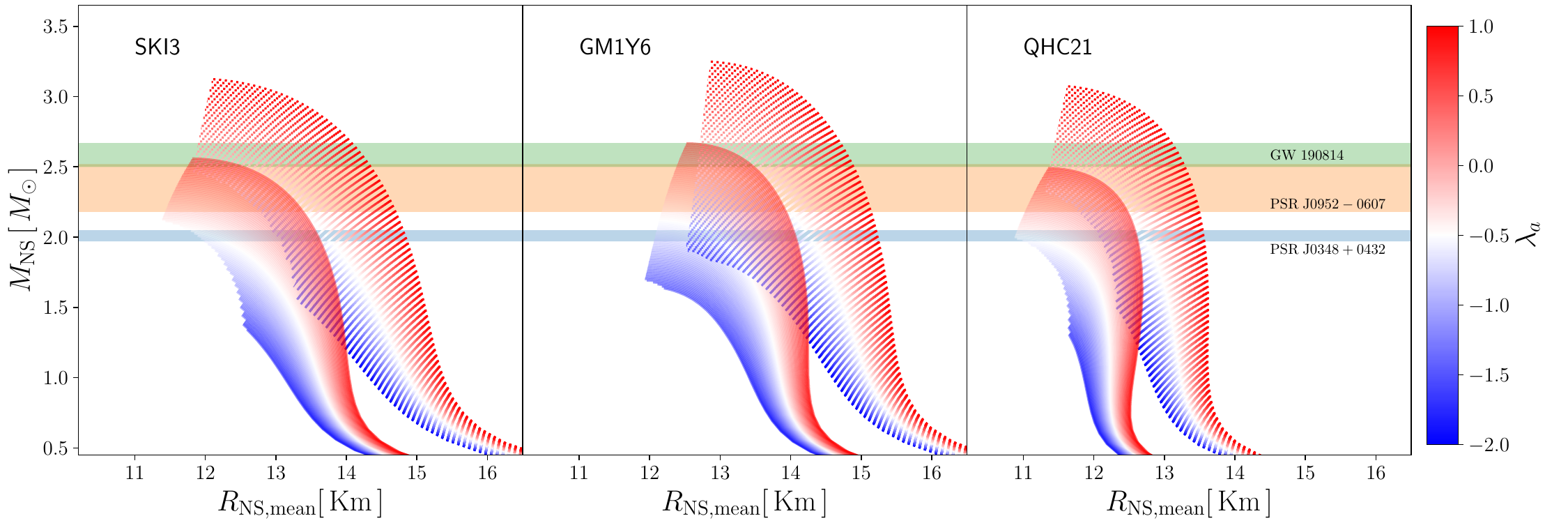}
    \caption{Gravitational mass and mean radius relation for anisotropic and slowly rotating stars and for three different NS EOS. The color scale corresponds to the value of the anisotropic parameter, $\lambda_a$. Solid lines represent static configurations, while dotted lines represent stars rotating at the Keplerian frequency. All the configurations shown here satisfy the causality conditions for the radial and tangential sound speed ($c_s<c$). The coloblack horizontal band indicates the observational constraints given by the pulsars PSR J0952-0607 \cite{2022ApJ...934L..17R}, PSR J0348-0432 \cite{2010Natur.467.1081D} and the mass of the secondary compact object of the GW190814 event \cite{2020ApJ...896L..44A}.}
    \label{fig:MRrot}
\end{figure*}
\begin{figure}
    \centering
    \includegraphics[width=0.98\columnwidth]{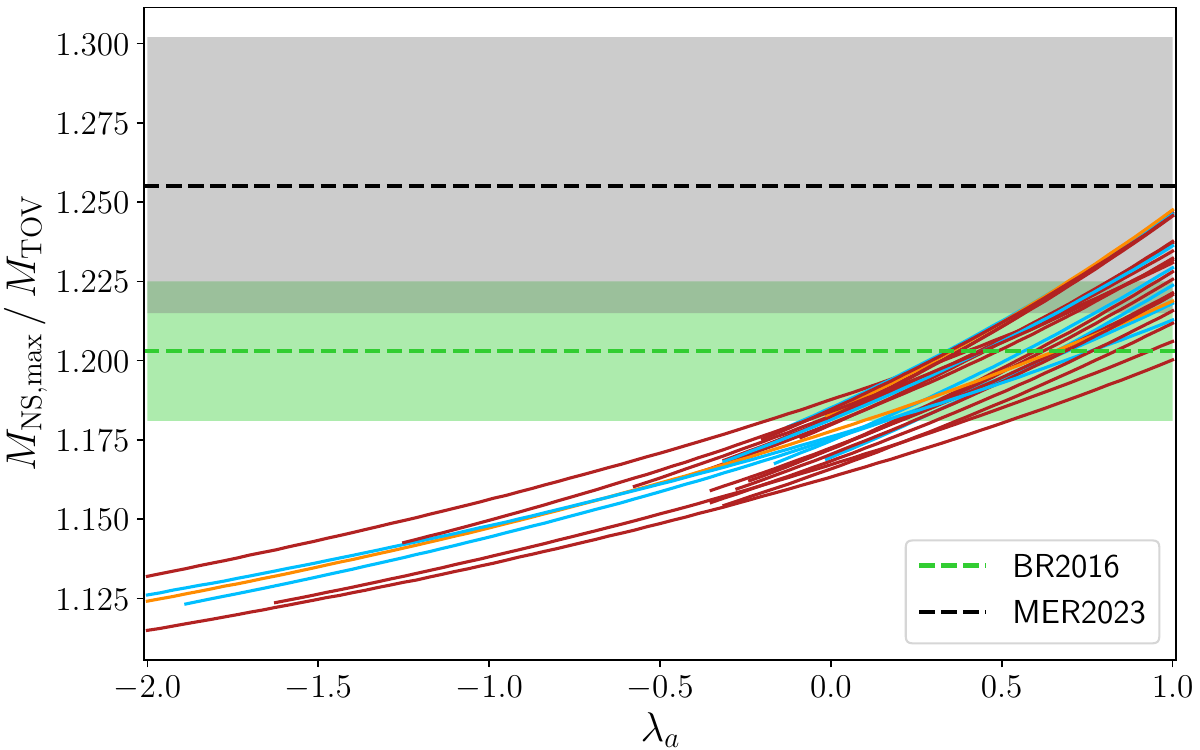}
    \caption{ Ratio between the maximum mass of the mass-shedding sequence, $M_{\rm NS,max}$, and the static sequence , $M_{\rm TOV}$, in function of the anisotropic parameter, $\lambda_a$. black, blue, and orange lines correspond to NS EOS that describes matter formed by nucleons, nucleons and hyperons, and nucleons, hyperons, and quarks, respectively. The coloblack bands represent the limits established by \citet{2016MNRAS.459..646B} (BR2016: $1.203\pm 0.022$) and   \citet{2024ApJ...962...61M} (MER2023: $1.255^{+0.047}_{-0.044}$)}
    \label{fig:Mmax}
\end{figure}

The total angular momentum of a rotating axisymmetric configuration is: 
\begin{equation}\label{eq:Jmoment}
    J = \int \sqrt{-g}\ T^t_\nu \xi^\nu_{(\phi)} \ \mathrm{d}^3 x = \int \sqrt{-g}\ T^t_\phi \ \mathrm{d}^3 x\, , 
\end{equation}
where $\xi^\nu_{(\phi)}$ is the killing vector associated with the axial symmetry and $g$ is the determinant of the metric. For the metric given by Eq.~(\ref{eq:ds2}), and using the Einstein's equations  up to $\mathcal{O}(\Omega^2)$, the total angular momentum of the star is:
\begin{equation}
   J = \frac{1}{6}R^4 \left( \frac{d\bar{\omega}}{dr}\right)_{R} + \mathcal{O}(\Omega^3).
\end{equation}
Solving equation (\ref{eq:omega}) outside the star (i.~e.~$\epsilon_0=P_0=\sigma_0 = 0$), it can be deduced that the angular velocity of the star is:
\begin{equation}
    \Omega = \bar{\omega}(R) + \frac{2J}{R^2}\ .
\end{equation}
The gravitational mass of the rotating configuration, $M_{\rm NS}$, is determined by the asymptotic behavior of the metric at large distances:
\begin{equation}
    g_{tt}\to 1 - \frac{2M_{\rm NS}}{r} \qquad {\rm for}\qquad r\to \infty
\end{equation}
Then, solving equations (\ref{eq:m0}) and (\ref{eq:h0}) outside the star and imposing regular conditions at $r=R$, it can be found:
\begin{equation}
    m_0(R) = \delta M - \frac{J^2}{R^3}\ ,
\end{equation}
with $\delta M = M_{\rm NS} - M(R)$, the difference between the gravitational mass of the rotating configuration and the corresponding static configuration with the same central density.

The surface of constant density of the rotating configuration is given by:
\begin{equation}
    R(r,\theta) = r + \xi_0(r) + \xi_2(r) P_2(\cos\theta),
\end{equation}
where
\begin{align}
     \xi_0 (R)&= \frac{p_0^\ast (P_0 + \epsilon)}{P_0 '} ,\\
     \xi_2 (R)&=  \frac{p_2^\ast (P_0 + \epsilon)}{P_0 '} . 
\end{align}
The proper circumferential radius of a circle on the surface around the axis of symmetry of the rotating configuration is:
\begin{equation}\label{eq:Rns}
\begin{aligned}
     R_{\rm NS}(\theta)\, = &\, \sqrt{g_{\phi\phi}(R,\theta)} \\
      \approx & \, R + \xi_0(R) + [\xi_2(R) + Rk_2(R)]P_2(\cos\theta) ,
\end{aligned}
\end{equation}
and the mean radius of the configuration is:
\begin{equation}\label{eq:Rnsmean}
     R_{\rm NS, mean} =  R + \xi_0(R)  .
\end{equation}
Figure~\ref{fig:MRrot}  shows the gravitational mass, $M_{\rm NS}$, and the mean radius, $R_{\rm NS, mean}$, relation for three representative NS equations of state: SKI3, GM1Y6, and QHC21. These EOS describe matter composed of nucleons, nucleons and hyperons, and nucleons, hyperons, and quarks, respectively. For each EOS and different values of the anisotropic parameter, $\lambda_a$, each panel of the Figure~\ref{fig:MRrot} shows two sets of configurations: the solid lines correspond to the static sequence ($\Omega=0$), while the dotted lines correspond to the mass shedding sequence. At this last limit, the angular velocity of the stars equals the angular velocity of a test particle on a circular orbit at the NS equator radius \cite[see][for the analytical expression of Keplerian frequencies for slow rotating stars]{2008AcA....58....1T,2013ApJ...762..117B}. As expected, rotation increases the gravitational mass and the radius of the star. The observational constraints given by PSR J0952-0607 \cite{2022ApJ...934L..17R} and PSR J0348-0432 \cite{2010Natur.467.1081D} are satisfied either by an anisotropic configuration or a uniformly rotating one,  in the cases where no static and isotropic stable configuration could do it. Objects in the mass gap, such as the secondary compact object of the GW190814 event \cite{2020ApJ...896L..44A}, might be explained by a combination of these two effects.

Figure~\ref{fig:MRrot} only shows configurations that satisfy the causality condition, meaning that the radial and tangential sound speeds do not exceed the speed of light inside the star. We consider that these configurations could potentially represent physical objects. For the slow rotation approximation, the tangential and radial sound speeds are defined as:
\begin{equation}
 c_{\rm s, r}^2 \equiv \frac{dP}{d\epsilon} \approx  \frac{dP_0}{d\epsilon_0}\left(1 - \frac{dP_0}{d\epsilon_0}\frac{d^2\epsilon_0}{dP_0^2}\delta P\right),
\end{equation}
\begin{equation}
 c_{\rm s, t}^2\equiv \frac{dP_\perp}{d\epsilon}\approx  \frac{dP_{\perp,0}}{d\epsilon_0} \left(1 - \frac{dP_0}{d\epsilon_0}\frac{d^2 \epsilon_0}{dP_0^2}\delta P\right).
\end{equation}

The Keplerian angular velocity is the maximum rate at which a star can rotate before it begins to shed mass. Consequently, the most massive rotating stars lie along the mass-shedding sequence. Figure~\ref{fig:Mmax} shows the ratio between the maximum mass of the mass-shedding sequence, $M_{\rm NS,max}$, and the maximum mass of the static sequence, $M_{\rm TOV}$, as a function of the anisotropic parameter, $\lambda_a$, for different NS EOS.
This ratio is an increasing function of the anisotropic parameter, always remaining in the range from 1.12 to 1.25. 
%This ratio is almost constant when the anisotropic parameter is negative and increases slightly when it becomes positive, always remaining in the range from 1.2 to 1.37. 

\citet{2016MNRAS.459..646B} provides a quasi-universal relation for the maximum mass of rotating stars as a function of the specific angular momentum through the study of 28 NS EOS, determining $M_{\rm NS,max}/M_{\rm TOV} \approx 1.203 \pm 0.022$. Later, using $10^6$ generic EOS, \citet{2024ApJ...962...61M} determined that the maximum-mass ratio between rotating and non-rotating stars is $1.255^{+0.047}_{-0.044}$. Figure~\ref{fig:Mmax} shows that this ratio applies to anisotropic rotating stars when $\lambda_a>0$. It is worth noting that these ratios were calculated when the corresponding static configuration with the maximum mass satisfies the causality condition for both the radial and tangential sound speeds. 

It is important to discuss the accuracy of the slow rotation regime. It requests that the rotation energy of the star, $T$, must be much smaller than its gravitational energy, $W$ \cite{1967ApJ...150.1005H}. By roughly estimating these quantities ($T\sim MR^2\Omega^2$ and $|W|\sim GM^2/R|$),  this condition traduces to: 

\begin{equation}
\Omega\ll \sqrt{\frac{GM}{R^3}} = \Omega_K^{J=0}.
\end{equation}

Our NS at the mass-shedding sequence rotates with $\Omega_K\sim (0.7-0.9)\Omega_K^{J=0}$, placing this sequence outside the slow rotation approximation. However,  \cite{1992ApJ...390..541W} shows that the mass of maximally rotating NSs derived with the HT formalism is accurate within an error of 4 \%, so we expect that the above result also applies in the fulling rotating case. Accurate limits on the maximum mass of rotating stars are important, for example,  in the analysis of binary NS mergers with electromagnetic counterparts \cite{2020MNRAS.499L..82M, 2018ApJ...852L..25R}.

%%%%%%%%%%%%%%%%%%%%%%%%%%%%%%%%%%%%%%%%%%%%
%%%%%%%%%%%%%%%%%%%%%%%%%%%%%%%%%%%%%%%%%%%%
\section{Universal Relations}\label{sec:UR}
%%%%%%%%%%%%%%%%%%%%%%%%%%%%%%%%%%%%%%%%%%%%
%%%%%%%%%%%%%%%%%%%%%%%%%%%%%%%%%%%%%%%%%%%%

In this section, to maintain consistency with the slow rotating approximation, we will consider configurations rotating with an angular velocity up to approximately $0.5 \Omega_K^{J=0}$ \cite{2005MNRAS.358..923B}. This guarantees a range of rotation frequencies that covers even the fastest pulsar observed, PSR J1748-2446ad, which has a rotational frequency of $716$~Hz \cite{2006Sci...311.1901H}.

%%%%%%%%%%%%%%%%%%%%%%%%%%%%%%%%%%%%%%%%%%%%########
\subsection{\label{subsec:IBE} Moment of Inertia   }

\begin{figure}
    \centering
    \includegraphics[width=0.99\columnwidth]{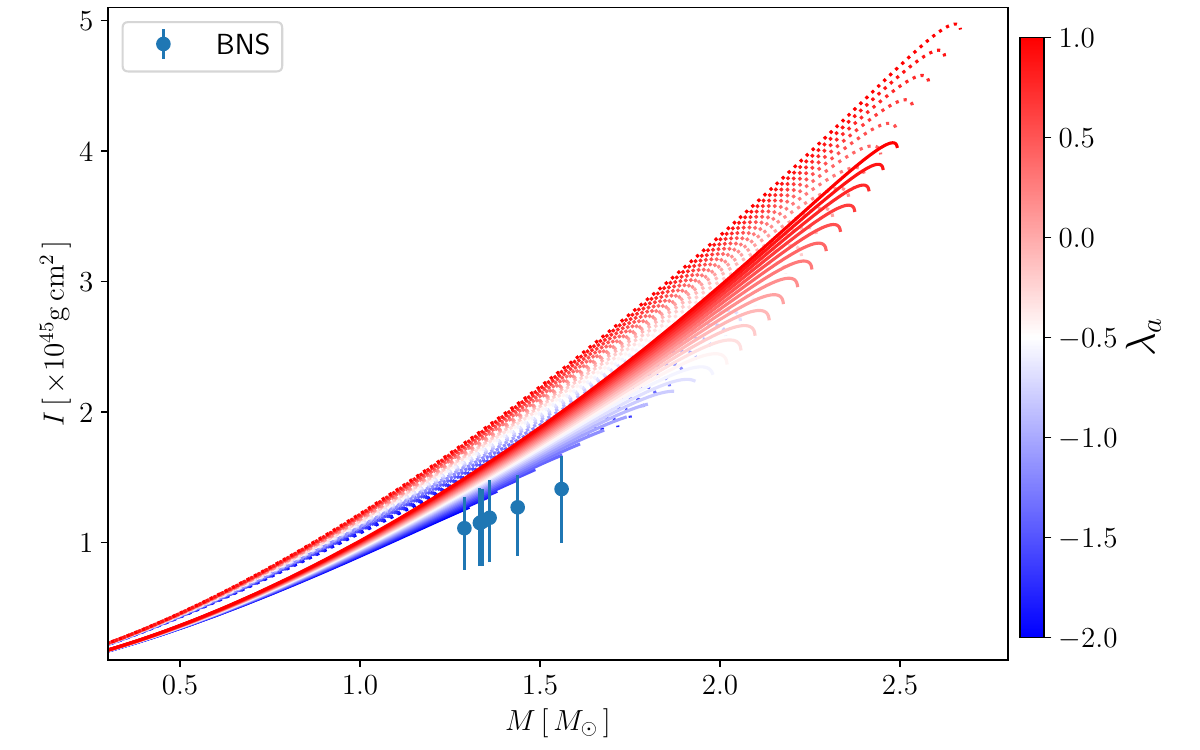}
    \caption{Moment of inertia, $I$, as a function of the non-rotating star mass, $M$, for two different EOS: GM1Y6 EOS (dotted lines) and QHC21 EOS (solid lines). The color scale corresponds to the value of the anisotropic parameter, $\lambda_a$. Dotted points correspond to the moment of inertia inferblack in \cite{2019PhRvD..99l3026K} for doubled pulsars. }
    \label{fig:Istar_M}
\end{figure}

The  moment of inertial of a star, $I$, with angular momentum, $J$, and angular velocity, $\Omega$, is:
\begin{equation}
    I\equiv\frac{J}{\Omega}\, .
\end{equation}
As can be deduced from equation~(\ref{eq:Jmoment}),  the moment of inertia is independent to the star's angular velocity up to $\mathcal{O}(\Omega^2)$.  Figure~\ref{fig:Istar_M} shows the moment of inertia as a function of the non-rotating gravitational mass for different values of the anisotropy parameter, $\lambda_a$, and for two NS EOS. Positive values of the anisotropy parameter increase the moment of inertia, while negative values decrease it. 

\begin{figure}
    \centering
    \includegraphics[width=0.99\columnwidth]{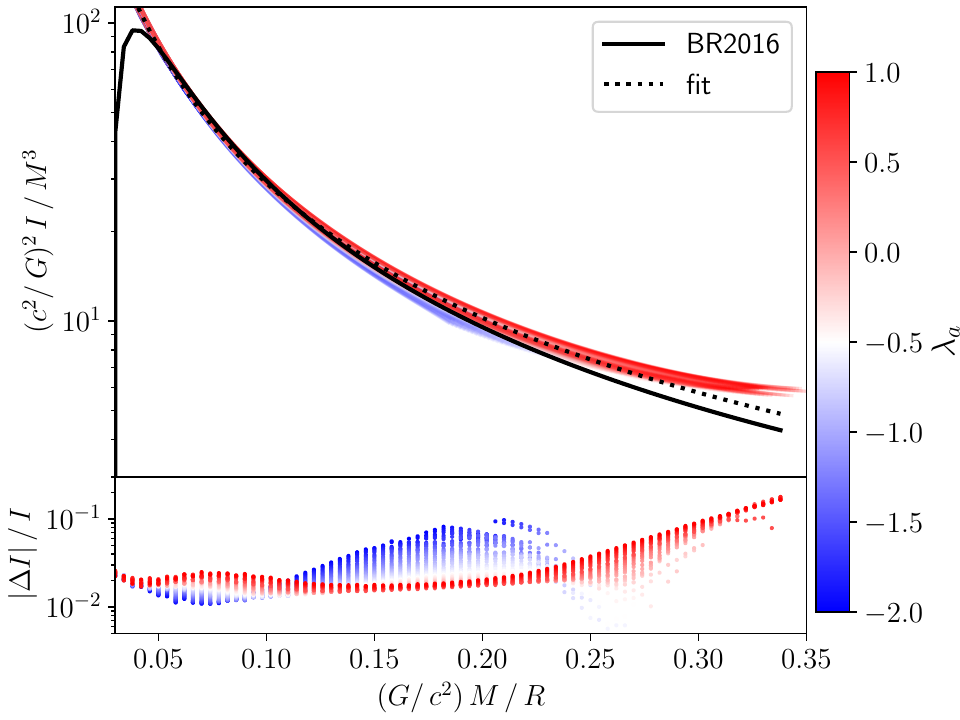}
    \caption{Normalized moment of inertia, $I/M^3$, as a function of the non-rotating star compactness, $M/R$, for different EOS. The color scale corresponds to the value of the anisotropic parameter, $\lambda_a$. The dotted black line corresponds to the fit given in equation~(\ref{eq:fit_I}). The bottom panel shows the relative difference between this fit and the numerical values of the normalized moment of inertia. We compare this fit with the one given by \citet{2016MNRAS.459..646B}  (solid black line).}
    \label{fig:Istar}
\end{figure}

Starting from the work of \citet{1994ApJ...424..846R}, many works have proposed universal relations for the NS moment of inertia. Following \citet{2016MNRAS.459..646B}, Figure~\ref{fig:Istar} shows the normalized moment of inertia, $(c^2/G)^2 I/M^3$, as a function of the compactness of the corresponding non-rotating configuration, $\mathcal{C}=GM/c^2 R$, for all the NS EOS of this work. As can be seen, the normalized moment of inertia is almost independent of the anisotropic parameter, $\lambda_a$, and seems to follow the universal relation given in \cite{2016MNRAS.459..646B} (solid black in Figure~\ref{fig:Istar}). We propose a slightly different fit relation that reproduces our results with an error of less than 10\%:
\begin{equation}\label{eq:fit_I}
    \frac{I}{M^3}\approx a_1 \mathcal{C}^{-1} + a_2 \mathcal{C}^{-2}  + a_3 \mathcal{C}^{-3}  + a_4 \mathcal{C}^{-4} ,
\end{equation}
with $a_1=1.019 \pm 0.004$, $a_2 = (2.253 \pm 0.005) \times 10^{-1}$, $a_3=(-3.871 \pm 0.014) \times 10^{-3}$ and $a_4=(2.328 \pm 0.001) \times 10^{-5}$. These kinds of universal relations are important because measuring the NS moment of inertia, along with its mass, directly constrains the NS EOS at high densities and its radius \cite[see][for some references]{2005ApJ...629..979L}. \cite{2020MNRAS.497.3118H} pblackicts that by 2030, the moment of inertia for double pulsars will be measublack with an accuracy of 11\%.

%
%%%%%%%%%%%%%%%%%%%%%%%%%%%%%%
\subsection{Binding-energy}
\begin{figure}
    \centering
    \includegraphics[width=0.99\columnwidth]{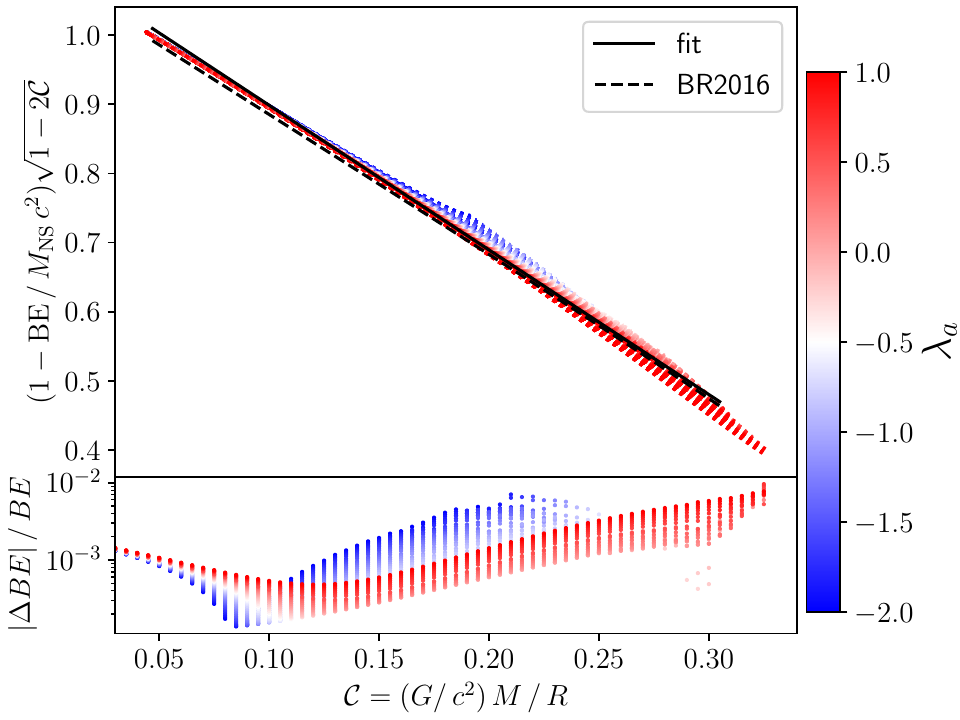}
    \caption{Normalized binding energy, $BE$  as function of the non-rotating star compactness, $M/R$. The color scale corresponds to the value of the anisotropic parameter, $\lambda_a$. The solid black line corresponds to the fit given in equation~(\ref{eq:fit_BE}). The bottom panel shows the relative difference between this fit and the numerical values star's binding energy. We compare this fit with the one given by \citet{2016MNRAS.459..646B}  (dashed black line). }
    \label{fig:BE}
\end{figure}

%%%%%%%%%%%%%%%%%%%%%%%%%%%%%%%
The baryonic mass for an axial symmetric  configuration is \cite{2003LRR.....6....3S}:
\begin{equation}
    M_B = \int \sqrt{-g}\,u^t\rho  \ \mathrm{d}^3x \, .
\end{equation}
By replacing the 4-velocity from equation~(\ref{eq:u_vec}), and expanding up to $\mathcal{O}(\Omega^2)$, the baryonic mass for slow rotating configurations is given by:
\begin{align}
    M_B = &\int_0^R  4\pi r^2 \rho \left(1-\frac{2M}{r}\right)^{-1/2}\left[ 1 + \frac{m_0}{r-2M} \right. \nonumber \\ & \left. +(\epsilon_0+P_0)p^*_0\frac{d \log(\rho)}{dP_0} +\frac{1}{3} e^{-\nu} r^2 \bar{\omega}^2 \right] dr \, .
\end{align}
The baryonic mass is affected by the monopolar deformation from rotation. Thus, the  binding energy of the star is:
\begin{equation}
    BE = ( M_B - M_{\rm NS} ) c^2\, .
\end{equation}
This energy can be understood as the amount of energy needed to form a stable configuration, encompassing both the gravitational and nuclear binding energy of the star. % or  the energy liberated during the birth of a NS. 
We propose the following relation for the NS biding energy for slow-rotating and anisotropic configurations:
\begin{equation}\label{eq:fit_BE}
  \left(  1-\frac{BE}{M_{\rm NS}c^2}\right)\sqrt{1-2\mathcal{C}}= c_1 +c_2 \mathcal{C}^2\, ,
\end{equation}
with $c_1=1.017\pm 0.001 $ and $c_2=-1.684\pm 0.001 $. Figure~\ref{fig:BE} shows the performance of the above fit for all the EOSs adopted in this paper and for different values for the anisotropic parameter, $\lambda_a$, obtaining an error of less than 1\%. The energy radiated via supernova neutrinos is a fundamental quantity closely linked to the gravitational binding energy of a NS \cite{2001ApJ...550..426L}. The core collapse of a massive star into a NS releases more energy if the resulting NS is more massive. Future detections of supernova explosions will provide valuable observational data to constrain the properties of NS matter \cite[see][for example]{1995ApJ...445L.129B, 2017JCAP...11..036G}.

\subsection{\label{subsec:Qstar} Quadrupole moment}
\begin{figure}
    \centering
    \includegraphics[width=0.99\columnwidth]{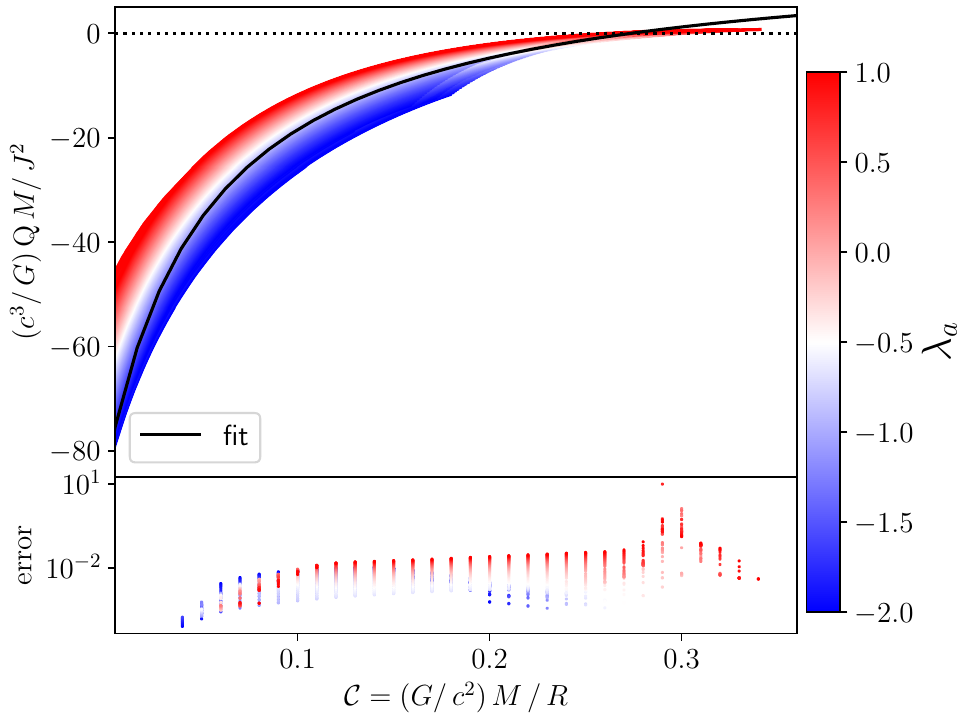}
    \caption{Normalize quadrupole moment as a function of the normalized angular momentum of slow rotating NSs. The color scale corresponds to the value of the anisotropic parameter, $\lambda_a$. The solid black line corresponds to the fit given in equation~(\ref{eq:fit_Q}), and the bottom panel shows the relative difference between this fit and the numerical values of the star's quadrupole moment. }
    \label{fig:QJ}
\end{figure}

An isotropic rotating NS becomes oblate, acquiring a negative quadrupole moment, $Q$. Following \cite{1999ApJ...512..282L}, the quadrupole moment of a slowly rotating star is defined as:
\begin{align}
    Q &= -\frac{5}{2}\lim_{r\to \infty} \left(\frac{r}{M}\right)^3\int_{-1}^{1} \log(H) P_2(\mu)d\mu \\
    &=  -\lim_{r\to \infty} \left(\frac{r}{M}\right)^3 h_2 = \frac{J^2}{M}+\frac{8}{5}BM^2. \nonumber
\end{align}
In the above definition, it has been used the solution of equations (\ref{eq:v2}) and (\ref{eq:h2}) outside the star \cite{1967ApJ...150.1005H}:
\begin{align}
 v_2 &=    -\frac{J^2}{r^4}+B \frac{2M}{r}\left(1-\frac{2M}{r}\right)^{-1/2}Q_2^1\left(\frac{r}{M}-1\right), \\
h_2 & = \frac{J^2}{r^4}\left(\frac{r}{M}+1\right)+ BQ_2^2\left(\frac{r}{M}-1\right),
\end{align}
with $Q_l^m$, the associated Legendre polynomials of the second kind, and $B$ a constant defined by the continuity condition through the surface of the functions $h_2$ and $v_2$. 

We propose a fit for the star's normalized quadrupole moment, $\bar{Q}$, and the star's compactness (for a Kerr black hole, $\bar{Q}=1$ \cite{1977MNRAS.179..483M}):
\begin{equation}\label{eq:fit_Q}
    \bar{Q}\equiv \frac{|Q|}{M^3}\left(\frac{M^2}{J}\right)^2 = b_1 + b_2\left(\frac{1}{\mathcal{C}}\right)^2 + b_3\left(\frac{1}{\mathcal{C}}\right)^3,
\end{equation}
with $b_1 =-4.638\pm 0.009$, $b_2=0.0028\pm 0.0001$ and $b_3=16.25 \pm 0.02$ (see Figure~\ref{fig:QJ}).  This adjustment depends on the star's anisotropy, with an error of less than 1\% for negative values of the anisotropic parameter, but an error of up to 10\% for positive values of the anisotropic parameter. Errors in the quadrupole moment could reach up to 20\% due to the slow-rotation approximation when $\Omega/\Omega_K^{J=0}> 0.1$, as indicated by \citet{2005MNRAS.358..923B}. It is worth saying that for highly anisotropic stars ($\lambda_a\approx 1$), the quadrupole moment could become positive, making the star prolate.

%%%%%%%%%%%%%%%%%%%%%%%%%%%%%%%%%%%%%%%%%%%%%%%%%%%%%%%%%%%%%%%%%%%%%%%%%%%%%%%%%%%%%%%%%%%

\section{\label{sec:discussion} Discussions and Conclusions}
%%%%%%%%%%%%%%%%%%%%%%%%%%%%%%%%%%%%%%%%%%%%%%%%%%%%%%%%%%%%%%%%%%%%%%%%%%%%%%%%%%%%%%%%%%%
In this paper, we model slowly rotating anisotropic  NS configurations by extending the HT formalism in general relativity \cite{1967ApJ...150.1005H,1968ApJ...153..807H}.  We derived the structure equations up to second order in the angular velocity, consistent with recent work \cite{2024PhRvD.110b4052B}.   To account for anisotropy,  we adopt the quasi-local relation for the tangential pressure proposed by \cite{PhysRevD.85.124023} (see Eq~\ref{eq:Delta_p}). Additionally, we study a wide range of nuclear EOS and parameterize them using the GPP formalism \cite{Boyle2020}.

We vary the anisotropic parameter $\lambda_a$, which controls the amount of anisotropy inside the star, from -2 to 1, following \cite{PhysRevD.109.043025}. We compute rotating anisotropic  NS configurations from zero angular velocity to the Keplerian angular velocity. Our results show that the ratio between the gravitational mass of the fastest rotating configurations and the corresponding static ones is  $1.12 \leq M_{\rm  NS, max }/M_{\rm TOV} \leq 1.25$, consistent with the results of \cite{2016MNRAS.459..646B,2024ApJ...962...61M}. The combination of anisotropy, in the form of equation~(\ref{eq:Delta_p}),  and rotation can increase the mass of a non-rotating isotropic NS configuration by up to 50\% \cite{PhysRevD.109.043025}. 

We proposed universal relations for the moment of inertia and the binding energy of rotating anisotropic  NS configurations with barotropic EOS as a function of its compactness (see equations~\ref{eq:fit_I} and \ref{eq:fit_BE}). For NS configurations rotating with $\Omega< 0.5 \Omega_K^{J=0}$, these relations are independent of the anisotropy parameter, $\lambda_a$, and are accurate with an error or less of 1 \%. Note that they may depend on the functional form of the anisotropy (equation~\ref{eq:Delta_p}). \textcolor{black}{\citet{2024arXiv240112519P} has also studied the universal relations for anisotropic quark stars, specifically between the star's moment of inertia, tidal deformability, and compactness. They also concluded that these relations are similarly insensitive to the equation of state and the presence of anisotropy.}

Finally, we compute the quadrupole moment of the slowly rotating anisotropic NS configurations.  We propose a universal relation involving the star's quadrupole moment and its compactness (see equation (\ref{eq:fit_Q})). In almost all cases, the quadrupole moment is negative, indicating that the star is oblate. However, when the anisotropy parameter $\lambda_a$ approaches unity, the star becomes prolate. This agrees with \cite{2024PhRvD.110b4052B}, who used the Bowers-Liang sphere and found that in the presence of high anisotropy, the star becomes prolate.  Note that our results may depend on the functional form of the anisotropy (equation~\ref{eq:Delta_p}). 

\begin{acknowledgments}
F.D.L-C is supported by the Vicerrectoría de Investigación y Extensión - Universidad Industrial de Santander, under Grant No. 3703. E.~A.~B-V is supported by the Vicerrector\'ia de Investigaci\'on y Extensi\'on - Universidad Industrial de Santander Postdoctoral Fellowship Program No. 2024000736. 
\end{acknowledgments}

%\appendix

\bibliography{apssamp}% Produces the 

\end{document}